\documentclass{nature}

\usepackage{amssymb}
\usepackage{amsmath}
\usepackage{wasysym}
\usepackage{graphicx}
\usepackage[usenames, dvipsnames]{color}
\usepackage[normalem]{ulem}
\usepackage{soul}

\usepackage[displaymath,mathlines]{lineno}

\usepackage{subfig}
\usepackage[font=footnotesize]{caption}

\usepackage{floatrow}
\DeclareFloatFont{tiny}{\scriptsize}
\title{Very high energy particle acceleration powered by the jets of the microquasar SS 433}

\newcounter{firstbib}

\begin{document}

\maketitle

\author{
A.U.~Abeysekara$^{1}$,
A.~Albert$^{2}$,
R.~Alfaro$^{3}$,
C.~Alvarez$^{4}$,
J.D.~\'{A}lvarez$^{5}$,
R.~Arceo$^{4}$,
J.C.~Arteaga-Vel\'{a}zquez$^{5}$,
D.~Avila Rojas$^{3}$,
H.A.~Ayala Solares$^{6}$,
E.~Belmont-Moreno$^{3}$
% ...
}

\begin{affiliations}
\small
  \item Department of Physics and Astronomy, University of Utah, Salt Lake City, UT, USA
  \item Physics and Theoretical Divisions, Los Alamos National Laboratory, Los Alamos, NM, USA
  \item Instituto de F\'{i}sica, Universidad Nacional Aut\'{o}noma de M\'{e}xico, Mexico City, Mexico
  \item Universidad Aut\'{o}noma de Chiapas, Tuxtla Guti\'{e}rrez, Chiapas, M\'{e}xico
  \item Universidad Michoacana de San Nicol\'{a}s de Hidalgo, Morelia, Mexico
  \item Department of Physics, Pennsylvania State University, University Park, PA, USA
  \item Department of Physics and Astronomy, University of Rochester, Rochester, NY, USA
  \item Department of Physics, Michigan Technological University, Houghton, MI, USA
  \item Instituto Nacional de Astrof\'{i}sica, \'{O}ptica y Electr\'{o}nica, Puebla, Mexico
  \item Institute of Nuclear Physics Polish Academy of Sciences, IFJ-PAN, Krakow, Poland
  \item Max-Planck Institute for Nuclear Physics, Heidelberg, Germany
  \item Facultad de Ciencias F\'{i}sico Matem\'{a}ticas, Benem\'{e}rita Universidad Aut\'{o}noma de Puebla, Puebla, Mexico 
  \item Departamento de F\'{i}sica, Centro Universitario de Ciencias Exactas e Ingenier\'{i}as, Universidad de Guadalajara, Guadalajara, Mexico 
  \item Department of Physics and Wisconsin IceCube Particle Astrophysics Center, University of Wisconsin-Madison, Madison, WI, USA 
  \item Instituto de Astronom\'{i}a, Universidad Nacional Aut\'{o}noma de M\'{e}xico, Mexico City, Mexico 
  \item School of Physics, Astronomy, and Computational Sciences, George Mason University, Fairfax, VA, USA 
  \item Department of Physics, University of Maryland, College Park, MD, USA
  \item Department of Astronomy, University of Maryland, College Park, MD, USA
  \item Joint Space-Science Institute, University of Maryland, College Park, MD, USA
  \item Inter-university Institute for High Energies, Universit\'{e} Libre de Bruxelles, Brussels, Belgium 
  \item NASA Marshall Space Flight Center, Astrophysics Office, Huntsville, AL, USA
  \item Department of Physics and Astronomy, University of New Mexico, Albuquerque, NM, USA
  \item Department of Physics and Astronomy, Michigan State University, East Lansing, MI, USA
  \item Universidad Politecnica de Pachuca, Pachuca, Mexico
  \item INFN and Universita di Padova, Padova, Italy
  \item Centro de Investigaci\'on en Computaci\'on, Instituto Polit\'ecnico Nacional, Mexico City, Mexico
  \item Universidad Aut\'{o}noma del Estado de Hidalgo, Pachuca, Mexico
  \item Instituto de Ciencias Nucleares, Universidad Nacional Aut\'{o}noma de Mexico, Mexico City, Mexico
  \item Santa Cruz Institute for Particle Physics, University of California, Santa Cruz, Santa Cruz, CA, USA
  \item School of Physics and Center for Relativistic Astrophysics, Georgia Institute of Technology, Atlanta, GA, USA
  \item Department of Physics, Stanford University, Stanford, CA, USA
  \item Department of Physics and Astronomy, University of California, Irvine, Irvine, CA, USA
  \item Physics Department, Centro de Investigacion y de Estudios Avanzados del IPN, Mexico City, Mexico
  \item Department of Physics and Astronomy, Purdue University, West Lafayette, IN, USA
\end{affiliations}

\begin{abstract}
SS 433 is a binary system containing a supergiant star that is overflowing its
Roche lobe with matter accreting onto a compact object (either a black hole or
neutron star)~\cite{Margon:1984anrev, Fabrika:2004asprv, Cherepashchuk:2005tg}.
Two jets of ionized matter with a bulk velocity of $\sim0.26c$ extend from the
binary, perpendicular to the line of sight, and terminate inside W50, a
supernova remnant that is being distorted by the jets~\cite{Zealey:1980mnras,
Margon:1989apj, Safi-Harb:1997apj, Eikenberry:2001tt, Migliari:2002tc,
Fabrika:2004asprv}. SS 433 differs from other microquasars in that the
accretion is believed to be super-Eddington~\cite{Mirabel:1999,
Begelman:2006mnras, Fabrika:2015erq}, and the luminosity of the system is
$\sim10^{40}$~erg~s$^{-1}$~\cite{Cherepashchuk:1982sa, Fabrika:2004asprv,
Mirabel:1999, Tetarenko:2015vrn}. The lobes of W50 in which the jets terminate,
about $40$~pc from the central source, are expected to accelerate charged
particles, and indeed radio and X-ray emission consistent with electron
synchrotron emission in a magnetic field have been
observed~\cite{Geldzahler:1980aa, Brinkmann:2006zt, Safi-Harb:1999apj}. At
higher energies ($>100$~GeV), the particle fluxes of $\gamma$ rays from X-ray
hotspots around SS~433 have been reported as flux upper
limits~\cite{Safi-Harb:1997apj, Aharonian:2005aa, Hayashi:2009ny,
Ahnen:2017tsc, Kar:2017wvr}. In this energy regime, it has been unclear whether
the emission is dominated by electrons that are interacting with photons from
the cosmic microwave background through inverse-Compton scattering or by
protons interacting with the ambient gas.  Here we report TeV $\gamma$-ray
observations of the SS~433/W50 system where the lobes are spatially resolved.
The TeV emission is localized to structures in the lobes, far from the center
of the system where the jets are formed. We have measured photon energies of at
least 25~TeV, and these are certainly not Doppler boosted, because of the
viewing geometry. We conclude that the emission from radio to TeV energies is
consistent with a single population of electrons with energies extending to at
least hundreds of TeV in a magnetic field of $\sim16$~micro-Gauss.
\end{abstract}

\begin{figure}[ht!]
  \includegraphics[width=0.6\textwidth]{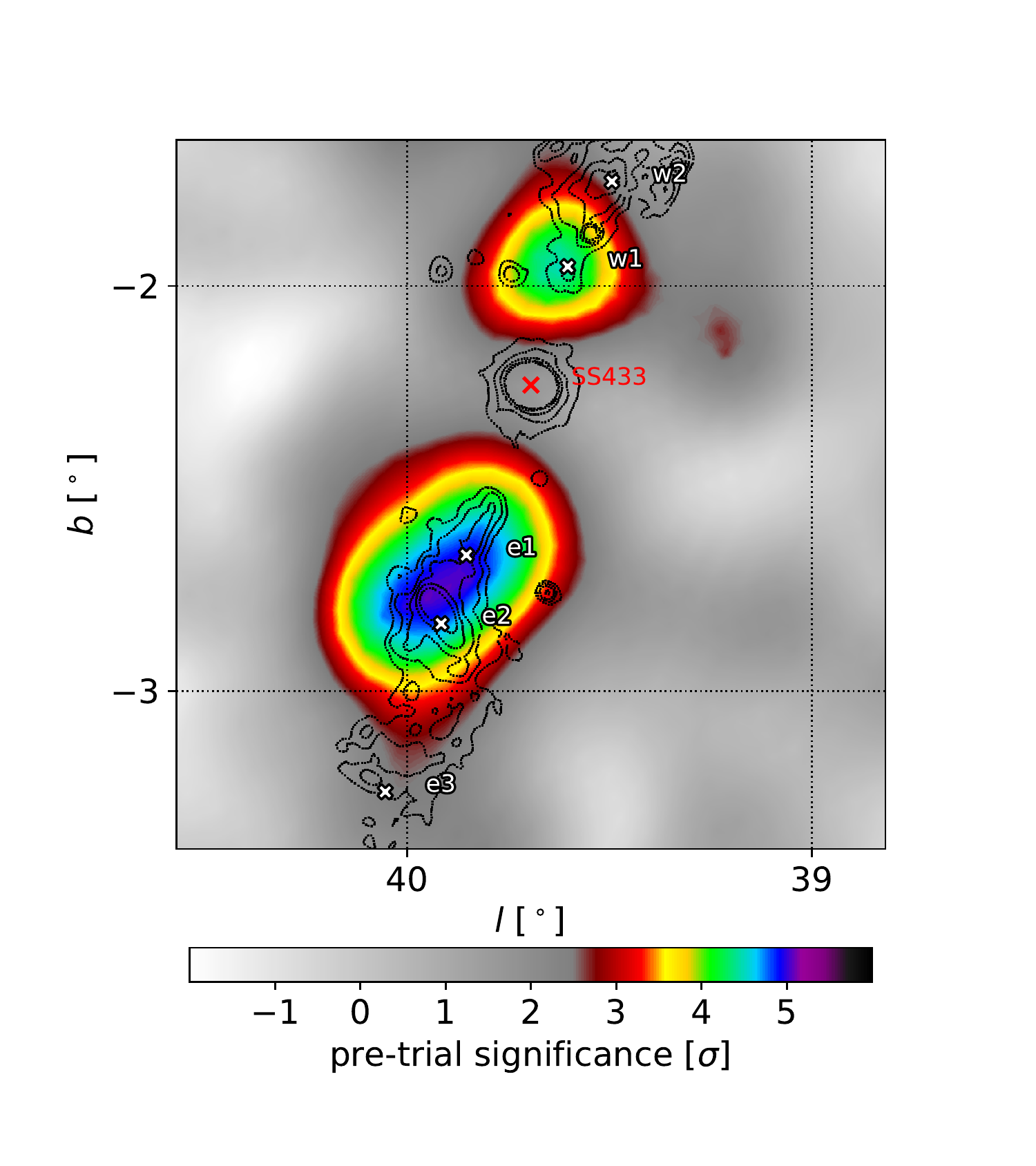}
  \caption{{\bf VHE $\gamma$-ray image of the SS~433/W50 region.} The color
  scale indicates the statistical significance of the excess counts above the
  background of nearly isotropic cosmic rays before accounting for statistical
  trials. The figure shows the $\gamma$-ray excess measured after the fitting
  and subtraction of $\gamma$-rays from the spatially extended source
  MGRO~J1908+06. The jet termination regions e1, e2, e3, w1, and w2 observed in
  the X-ray data are indicated, as well as the location of the central binary.
  The solid contours show the X-ray emission observed from this system.}
  \label{fig:ss433_obs}
\end{figure}


In the SS~433/W50 complex, several regions located west of the central binary
(w1 and w2) and east (e1, e2, e3) are observed to emit hard
X-rays~\cite{Safi-Harb:1997apj}. Previous searches for very high-energy (VHE)
$\gamma$-ray emission from the hotspots between roughly $100$~GeV and $10$~TeV
have produced null results~\cite{Aharonian:2005aa, Hayashi:2009ny,
Ahnen:2017tsc, Kar:2017wvr}, though an excess observed at $\sim800$~MeV may be
associated with SS~433 and W50 \cite{Bordas:2014kla}.  The High Altitude Water
Cherenkov (HAWC) observatory is a wide field-of-view VHE $\gamma$-ray
observatory surveying the Northern sky above 1~TeV, and is optimized for photon
detection above 10~TeV~\cite{Abeysekara:2017hyn}. SS~433 transits $15^\circ$
from the zenith of the HAWC detector each day, and has been observed with
$>90\%$ uptime since the start of detector operations in 2015.

In 1017 days of measurements with HAWC, an excess of $\gamma$ rays with a
post-trials significance of $5.4\sigma$ has been observed in a joint fit of the
eastern and western interaction regions of the jets of SS~433.  The emission is
plotted in galactic coordinates in Fig.~\ref{fig:ss433_obs}, which includes an
overlay of the X-ray observations of the jets and the central binary.  The
$\gamma$-ray emission is spatially coincident with the X-ray hotspots w1 and
e1; no significant emission is observed at the location of the central binary
where the jets are produced.

Spatial and spectral fits to SS~433 are performed in a semicircular region of
interest (RoI) designed to mask out diffuse emission from the Galactic Plane.
The RoI also removes significant spatially extended emission from the nearby
$\gamma$-ray source MGRO~J1908+06. The spatial distribution and spectrum of
$\gamma$ rays from MGRO~J1908+06 are fit using an electron diffusion
model~\cite{Lopez-Coto:2017uku}, and point-like sources centered on e1 and w1
are fit on top of this extended emission. As a systematic check, the regions
are also fit using X-ray spatial templates and extended Gaussian functions.
Neither improves the statistical significance of the fits. Upper limits on the
angular size of the emission regions are $0.25^\circ$ for the east hotspot and
$0.35^\circ$ for the west hotspot at 90\% confidence. Given the distance to the
source of $5.5$~kpc, this corresponds to a physical size of $24$~pc and
$34$~pc, respectively. The constraint is tighter on the eastern hotspot due to
its higher statistical significance.

The VHE $\gamma$-ray flux is consistent with a hard $E^{-2}$ spectrum, though
current data from HAWC are not of sufficient significance to constrain the
spectral index.  Therefore we report the flux of both hotspots at 20~TeV, where
systematic uncertainties due to the choice of spectral model are minimized and
the sensitivity of HAWC is maximized. At e1, the VHE flux is
$2.4^{+0.6}_{-0.5}(\text{stat.})^{+1.3}_{-1.3}(\text{syst.})\times10^{-16}$
TeV$^{-1}$~cm$^{-2}$~s$^{-1}$, and at w1 the flux is
$2.1^{+0.6}_{-0.5}(\text{stat.})^{+1.2}_{-1.2}(\text{syst.})\times10^{-16}$
TeV$^{-1}$~cm$^{-2}$~s$^{-1}$. HAWC detects $\gamma$ rays from the interaction
regions up to at least 25~TeV. The energies of these $\gamma$
rays are a factor of three to ten higher than previous measurements from
microquasars~\cite{Albert:2006vk,Archambault:2016nmo}. Since most
$\gamma$-ray telescopes are optimized for measurements below 10~TeV, this may
explain why these photons were not observed in previous observational
campaigns.

The $\gamma$ rays detected by HAWC are produced by radiative or decay processes
from much higher energy particles. The detection yields important information
about the mechanisms and sites of particle acceleration, the types of particles
accelerated (e.g., protons or electrons), and the radiative processes which
produce the spectrum of emission from radio to VHE $\gamma$ rays.  Two
scenarios for explaining the HAWC observations of the e1 and w1 regions can be
tested. The first is that protons are primarily responsible for the observed
$\gamma$ rays. Protons must have an energy of at least $250$~TeV to produce
$25$~TeV $\gamma$ rays through hadronic collisions with ambient gas.
Proton-proton collisions yield $\pi^0$ particles which decay to VHE $\gamma$
rays, and $\pi^\pm$ particles which decay to the secondary electrons and
positrons responsible for radio to X-ray emission via synchrotron radiation.
This scenario is of particular interest because there is spectroscopic evidence
for ionized nuclei in the inner jets of SS~433~\cite{Migliari:2002tc,
Reynoso:2008nk}. The alternative scenario requires electrons of at least
$130$~TeV to up-scatter the low-energy photons from the cosmic microwave
background (CMB) to $25$~TeV $\gamma$ rays. In this case, the radio to X-ray
emission is dominated by synchrotron radiation from this same population of
electrons in the magnetized plasma of the jets and lobes.

The fact that the VHE emission is detected along a line of sight nearly
orthogonal to the jet axis means that charged particle trajectories become
isotropic before they interact to produce the $\gamma$ rays. The embedded
magnetic fields in the VHE regions can easily deflect the accelerated particles
because their typical gyroradii are much smaller than the size of the emission
regions, $\sim30$~pc. The jets are only mildly relativistic, so the emission
from the interaction regions will have a negligible Doppler beaming effect and
remain nearly isotropic.

\begin{figure}[ht!]
  \includegraphics[width=0.9\textwidth]{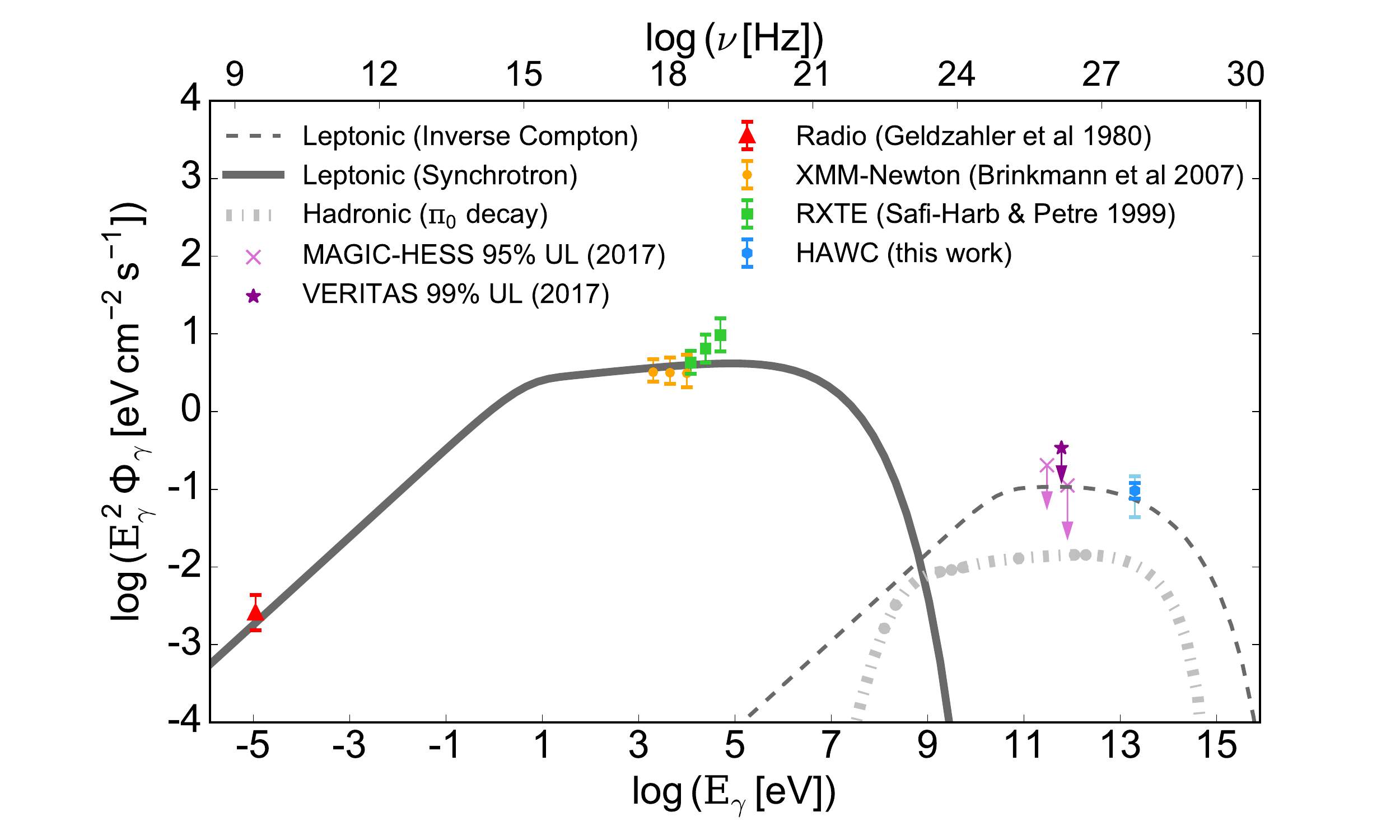}
  \caption{{\bf Broadband spectral energy distribution of the eastern emission
  region.} The data include radio~\cite{Geldzahler:1980aa}, soft
  X-ray~\cite{Brinkmann:2006zt}, hard X-ray~\cite{Safi-Harb:1999apj}, and VHE
  $\gamma$-ray upper limits~\cite{Ahnen:2017tsc, Kar:2017wvr}, and HAWC
  observations of e1. Error bars indicate $1\sigma$ uncertainties, with the
  thick (thin) errors on the HAWC flux indicating statistical (systematic)
  uncertainties and arrows indicating flux upper limits.  The multiwavelength
  spectrum produced by electrons assumes a single electron population following
  a power-law spectrum with an exponential cutoff.  The electrons produce radio
  to X-ray photons through synchrotron emission in a magnetic field (thick
  solid line) and TeV $\gamma$ rays through inverse Compton scattering of the
  cosmic microwave background (thin dashed line). The dash-dotted line
  represents the radiation produced by protons, assuming 10\% of the jet
  kinetic energy converts into protons.}
  \label{fig:ss433_model}
\end{figure}

The flux of VHE $\gamma$ rays observed by HAWC makes the proton scenario for
SS~433 unlikely, because the total energy required to produce the highly
relativistic protons is too high. The jets of SS~433 are known to be
radiatively inefficient, with most of the jet energy transformed into the
thermal energy of W50~\cite{Safi-Harb:1999apj,Panferov:2016rcw} rather than
particle acceleration. We model the primary proton spectrum as a power law with
an exponential cutoff, $dN/dE_p\propto E_p^{-2} \exp{(-E_p/1~\text{PeV})}$. If
we assume 10\% of the jet kinetic energy converts into accelerated protons, and
the ambient gas density is
$0.05$~cm$^{-3}$~\cite{Safi-Harb:1999apj,Panferov:2016rcw}, then the resulting
flux of $\gamma$ rays from proton-proton collisions is much less than the
observed $\gamma$-ray flux, as shown in the dash-dotted line of
Fig.~\ref{fig:ss433_model}. In fact, for a target proton density as large as
$0.1$~cm$^{-3}$ in the e1 region \cite{Safi-Harb:1999apj,Panferov:2016rcw}, the
total energy of the proton population needs to be $\sim3\times10^{50}$~erg to
explain the observed $\gamma$ rays, assuming an $E_{\gamma}^{-2}$ spectrum.
This is comparable to the total jet energy available during the presumed
30,000-year lifetime of SS~433~\cite{Fabrika:2004asprv}. Furthermore, because
the synchrotron emission from secondary electrons from charged pion decay is
always lower than the $\gamma$-ray flux from $\pi^0$ decay, and the observed
X-ray flux is higher than the $\gamma$-ray flux, the X-rays cannot originate
solely from secondary electrons.  Finally, the proton scenario requires that
the protons remain trapped in the region observed by HAWC for the lifetime of
SS~433~\cite{Fabrika:2004asprv}. This means the protons must diffuse very
slowly, with a diffusion coefficient $\sim1/1000$ of the typical value of the
interstellar medium (ISM)
$D_\text{ISM}\approx3\times10^{28}(E/3~\text{GeV})^{1/3}$~cm$^2$~s$^{-1}$
\cite{Ptuskin:2005ax}. This value, comparable to the theoretical Bohm limit, is
very small but not impossible.  Given the uncertainties in the historical jet
flux, the ambient particle density, and the radiative efficiency, we cannot
exclude the possibility that some fraction of the $\gamma$-ray flux is produced
by protons.  However, we do rule out the possibility that the VHE $\gamma$ rays
are entirely produced by protons.

Highly relativistic electrons, on the other hand, can produce $\gamma$ rays
much more efficiently, primarily via inverse Compton scattering of CMB photons
to $\gamma$ rays.  The inverse Compton losses due to up-scattering of infrared
and optical photons are suppressed due to the Klein-Nishina effect and are thus
dominated by scattering of CMB photons~\cite{Moderski:2005jw}. In this
scenario, the ratio of the VHE $\gamma$-ray to X-ray fluxes constrains the
energy density in the magnetic field compared to the energy density in CMB
photons.  We have modeled the broadband spectral energy distribution (SED) of
the eastern emission region $15'$ to $33'$ from the center of SS~433.  The
solid and dashed lines in Fig.~\ref{fig:ss433_model} show the SED of a leptonic
model for e1 produced by an injected flux of relativistic electrons with an
energy spectrum $dN/dE\propto E^{-\alpha}\exp{(-E/E_\text{max})}$ in a magnetic
field of strength $B$. We use the parameters $\alpha=1.9$,
$E_\text{max}=3.5$~PeV, and $B=16~\mu$G (see the methods section). The estimate
of the magnetic field strength is consistent with the equipartition of energy
between the relativistic electrons and magnetic fields, which is common in
astrophysical systems \cite{Safi-Harb:1999apj}. The required total energy
budget for relativistic electrons is three orders of magnitude lower than the
total jet energy.

The maximum electron energy of $\sim1$~PeV has significant implications for
electron acceleration sites and acceleration mechanisms in SS~433. SS~433 is
distinguished from other binary systems with relativistic objects because it
realizes a supercritical accretion of gas onto the central
engine~\cite{Fabrika:2004asprv}. Powerful accretion flows and the inner jets
near the compact object have therefore been proposed as possible acceleration
sites of relativistic particles~\cite{Reynoso:2008nk}. However, the observation
from HAWC suggests that ultra-relativistic electrons are not accelerated near
the center of the binary. If electrons were accelerated in the central region,
they would have cooled by the time they reached the sites of observed VHE
emission. Due to their small gyroradii, high-energy electrons may transport in
a magnetized medium via diffusion or advection.  The distance traveled via
diffusion within the cooling time of an electron of energy $E$ is
$
  r_d = 2\sqrt{Dt_\text{cool}}\approx36~\text{pc}\
        (E/1~\text{PeV})^{-1/3} (B/16~\mu\text{G})^{-1}
$,
using the diffusion coefficient typical of the ISM~\cite{Ptuskin:2005ax}.
This distance would be even smaller for diffusion coefficients lower than the
typical ISM value. Similarly, the distance traveled by electrons being advected
with the jet flow is
$
  r_\text{adv} = 0.26c\cdot t_\text{cool}\approx4~\text{pc}\
    (E/1~\text{PeV})^{-1} (B/16~\mu\text{G})^{-2}
$
for a jet velocity of $0.26c$. Both distance scales are smaller than the
$40$~pc distance between the binary and e1, indicating the electrons are
not accelerated near the center of the system.

Instead, the highly energetic electrons in SS~433 are likely accelerated in the
jets and near the VHE $\gamma$-ray emission regions. This presents a challenge
to current acceleration models.  For example, particle acceleration may be
driven by the dissipation of the magnetic fields in the jets, but above several
hundred TeV the electron acceleration time exceeds the electron cooling time,
assuming a $16~\mu\text{G}$ magnetic field. Thus the system does not appear to
have sufficient acceleration power, unless there are very concentrated magnetic
fields along the jets. If instead particle acceleration is driven by standing
shocks produced by the bulk flow of the jets, it is possible to reach PeV
energies if the size of the acceleration region is larger than the electrons'
gyroradii.  However, shocks in the interaction regions are not currently
resolved by X-ray or $\gamma$-ray measurements.

Studies of microquasars such as SS~433 provide valuable probes of the particle
acceleration mechanisms in jets, since these objects are believed to be scale
models of the much larger and more powerful jets in active galactic nuclei
(AGN) \cite{Romero:2016hjn}. AGN are the most prevalent VHE extragalactic
sources and are believed to be the sources of the highest energy hadronic
cosmic rays. While AGN are not spatially resolved at VHE energies, with this
observation we have identified a VHE source in which we can image particle
acceleration powered by jets. Future high-resolution VHE observations of SS~433
are possible with pointed atmospheric Cherenkov telescopes to better localize
the emission sites, and further high-energy measurements with HAWC will record
the spectrum at high energies and better constrain the maximum energy of
accelerated particles.

\newpage

{\bf References}
\vspace{1em}




\newpage

\begin{addendum}

  \item We acknowledge the support from: the US National Science Foundation
  (NSF); the US Department of Energy Office of High-Energy Physics; the
  Laboratory Directed Research and Development program of Los Alamos National
  Laboratory; Consejo Nacional de Ciencia y Tecnolog\'{i}a, M\'{e}xico (grants
  271051, 232656, 260378, 179588, 239762, 254964, 271737, 258865, 243290,
  132197, and 281653) (C\'{a}tedras 873, 1563); Laboratorio Nacional HAWC de
  rayos gamma; L'OREAL Fellowship for Women in Science 2014; Red HAWC,
  M\'{e}xico; DGAPA-UNAM (Direcci\'{o}n General Asuntos del Personal
  Acad\'{e}mico-Universidad Nacional Aut\'{o}noma de M\'{e}xico; grants
  IG100317, IN111315, IN111716-3, IA102715, IN109916, IA102917); VIEP-BUAP
  (Vicerrector\'{i}a de Investigaci\'{o}n y Estudios de Posgrado-Benem\'{e}rita
  Universidad Aut\'{o}noma de Puebla); PIFI (Programa Integral de
  Fortalecimiento Institucional) 2012 and 2013; PROFOCIE (Programa de
  Fortalecimiento de la Calidad en Instituciones Educativas) 2014 and 2015; the
  University of Wisconsin Alumni Research Foundation; the Institute of
  Geophysics, Planetary Physics, and Signatures at Los Alamos National
  Laboratory; Polish Science Centre grant DEC-2014/13/B/ST9/945 and
  DEC-2017/27/B/ST9/02272; and Coordinaci\'{o}n de la Investigaci\'{o}n
  Cient\'{i}fica de la Universidad Michoacana.  Thanks to S. Delay, L.
  D\'{i}az, and E. Murrieta for technical support. We acknowledge Richard
  Mushotzky for providing the spectrum of the XMM-Newton data in the HAWC
  detection region.

  \item[Author Contributions] C.~D. Rho (crho2@ur.rochester.edu) and H. Zhou
  (hao@lanl.gov) analyzed the data and performed the maximum likelihood
  analysis. Modeling of the leptonic and hadronic emission has been carried out
  by K.~Fang (kefang@umd.edu) and H.~Zhang (zhan3038@purdue.edu). S.~BenZvi
  (sybenzvi@pas.rochester.edu) and B.~Dingus (dingus@lanl.gov) helped prepare
  the manuscript. The full HAWC Collaboration has contributed through the
  construction, calibration, and operation of the detector, the development and
  maintenance of reconstruction and analysis software, and vetting of the
  analysis presented in this manuscript. All authors have reviewed, discussed,
  and commented on the results and the manuscript.

  \item[Competing Interests] The authors declare no competing interests.

  \item[Data Availability] The datasets analyzed during this study are
  available at a public data repository maintained by the HAWC Collaboration:
  https://data.hawc-observatory.org/.

  \item[Code Availability] The study was carried out using the Analysis and
  Event Reconstruction Integrated Environment Likelihood Fitting Framework
  (AERIE-LiFF) and the Multi-Mission Maximum Likelihood (3ML) software.
  The code is open-source and publicly available on Github:
  https://github.com/rjlauer/aerie-liff and https://github.com/giacomov/3ML.
  The software includes instructions on installation and usage.

\end{addendum}

\newpage
\begin{center}
  \bf{Methods}
\end{center}

\renewcommand\tablename{Extended Data Table}
\renewcommand\figurename{Extended Data Figure}
\setcounter{figure}{0}

\subsection{Data Reduction and Maximum Likelihood Analysis}

This analysis uses 1017 days of data from the High Altitude Water Cherenkov
Observatory collected between November 26, 2014 and December 20, 2017. HAWC is
an array of 300 tightly packed identical water Cherenkov detectors deployed
4100~m above sea level on the slope of volcano Sierra Negra,
Mexico~\cite{Smith:2015wva}. Each detector is a cylindrical water tank standing
5~m tall and 7.3~m in diameter, filled with 180~000~L of purified water. At the
bottom of each tank are four photomultipliers (PMTs) facing upward. The PMTs
record the Cherenkov photons created by the relativistic secondary particles
produced when primary cosmic rays and $\gamma$ rays interact at the top of the
atmosphere. The HAWC array covers 22,000~m$^2$. Its construction ended in
December 2014, and the full array was commissioned in March 2015.

Using the relative arrival time of photoelectrons (``hits'') detected by the
PMTs, the arrival direction of primary $\gamma$ rays can be reconstructed with
an accuracy of $\sim1^\circ$ below 1~TeV to $<0.2^\circ$ above
10~TeV~\cite{Abeysekara:2017mjj}. The accuracy of the reconstruction determines
the point spread function (PSF) of the detector, and is a function of the
energy, zenith angle, and composition of the primary particle. Air showers from
$\gamma$ rays are discriminated from the cosmic-ray background by filtering out
``clumpy'' patterns of hits, which are characteristic of the energy deposited
by hadronic cosmic rays. The cosmic-ray background rejection efficiency
improves rapidly as a function of energy, increasing from $90\%$ at 1~TeV to
$99.9\%$ at 10~TeV~\cite{Abeysekara:2017mjj}.

To compute the statistical significance of $\gamma$-ray emission observed with
HAWC, a maximum likelihood fit using parametric spatial and spectral models is
applied to the data~\cite{Younk:2015saa,Vianello:2015tuw}. The models are
forward-folded through the detector response to produce expected counts of
$\gamma$-ray signal events and cosmic-ray background events. The expectation is
then compared to the observed counts $N_\text{obs}$.  To calculate the expected
counts as a function of position on the sky, the events are binned in a fine
mesh using the HEALPix pixelization of the unit sphere~\cite{Gorski:2004by}.
The pixelization is chosen to be $0.1^\circ$, roughly two to five times smaller
than the radius of the instrument PSF. To apply models of the energy spectrum
of a source, the data are binned according to the fraction of PMTs in the
detector triggered by an air shower~\cite{Abeysekara:2017mjj}. This measure of
shower ``size'' is used as a coarse proxy for the energy of the primary
particle; a total of 9 size bins $B$ is used.

Given a model with $\vec{\theta}$ spatial and spectral parameters, the maximum
likelihood of the model having produced the data is
\begin{equation}\label{eq:ml}
  \ln{\mathcal{L}(N_\text{obs}|\vec{\theta})}
    = \sum_{B=1}^9\sum_{j=1}^m
    \ln{P(N_\text{obs}^{j,B}|\vec{\theta})},
\end{equation}
where the sum runs over the size bins $B$ and the HEALPix pixels $j$ in the
region of interest (RoI) of the fit. $P$ is the Poisson probability of
detecting $N_\text{obs}^{j,B}$ events in pixel $j$ and size bin $B$ given the
model parameters $\vec{\theta}$.

Within the RoI around SS~433 defined in Extended Data Fig.~\ref{fig:ss433_obs},
two fits are performed to maximize the likelihood: a fit which only accounts
for the emission from MGRO J1908+06 (null hypothesis), and the combined
emission from MGRO J1908+06 and the SS~433 region (alternative hypothesis). The
ratio of the maximum likelihood defines a test statistic
\begin{equation}\label{eq:ts} \text{TS} =
2\left(\ln{\mathcal{L}(N_\text{obs}|\vec{\theta}_\text{alt})} -
\ln{\mathcal{L}(N_\text{obs}|\vec{\theta}_0)}\right), \end{equation}
where $\vec{\theta}_0$ and $\vec{\theta}_\text{alt}$ represent the spatial and
spectral parameters of the null and alternative hypotheses, respectively. TS is
then converted to a $p$-value to estimate the statistical significance of
emission from SS~433. As discussed in the main text, the alternative hypothesis
assumes two point sources with power law spectra $dN/dE=f_0\cdot
(E/20~\text{TeV})^{-2}$, where the flux normalization $f_0$ is the free
parameter of the spectral model.

Wilks' Theorem is used to convert TS to a $p$-value~\cite{Wilks:1938}. In the
joint likelihood maximization, there are 2 degrees of freedom for the two
separately fit flux normalizations of the hotspots at w1 and e1. Therefore, we
calculate the one-tailed $p$-value
$\text{pr}(\text{TS}>\chi^2=41.2|\text{dof}=2)=1.13\times10^{-9}$. Since the
positions of the point source fits at w1 and e1 were chosen after looking into
the data, and because we are searching for other microquasars in the field of
view of HAWC, we must apply {\sl a posteriori} corrections to the $p$-value to
account for multiple-comparison effects.

The X-ray interaction regions w1, w2, e1, e2, and e3 are {\sl a priori}
candidates for the locations of the maxima, as is the center of the binary
system, for a total of six potential hotspots. Given the angular resolution of
HAWC, it would not be possible to spatially resolve all six hotspots; at best
three regions (east, west, and center) can be separately fit with confidence.
There are 23 possible combinations of the six {\sl a priori} locations which
can be used to fit one, two, or three hotspots in the eastern, central, and
western regions of the source. We add an additional 12 trials to account for
the known microquasars in the field of view of HAWC~\cite{Mirabel:1999,
Chaty:2006, Tetarenko:2015vrn}. This trial factor is conservative given that
several galactic microquasars are already known TeV
sources~\cite{Aharonian:2005sci, Aliu:2013rpa}.

Given 35 total trials, the corrected $p$-value is $3.96\times10^{-8}$, which
corresponds to a statistical significance of $5.4\sigma$.

\subsection{Modeling of the Nearby Extended Source MGRO~J1908+06}

\begin{figure}[ht!]
  \includegraphics[width=\textwidth]{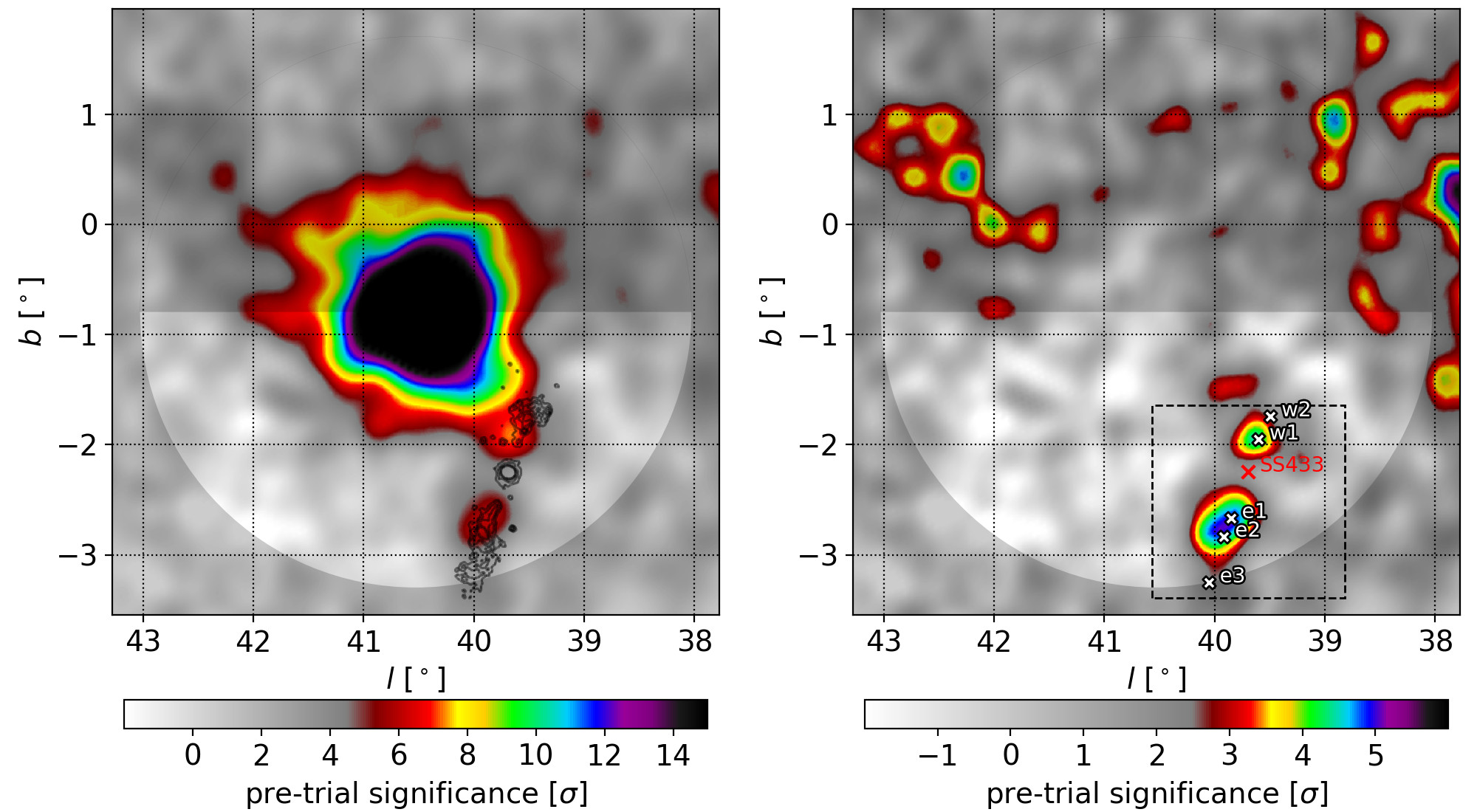}
  \caption{{\bf VHE $\gamma$ rays from MGRO~J1908+06 and SS~433/W50.}
  The color scale indicates the statistical significance of the excess counts
  above background before accounting for statistical trials.
  {\sl Left panel}: the bright extended $\gamma$-ray source MGRO~J1908+06 is shown at the
  center of the image and SS~433/W50 at the bottom. The dark contours show
  X-ray emission from SS~433 and its jets~\cite{Brinkmann:1996aa}. The
  semicircular area indicates the
  region of interest used to fit the $\gamma$-ray observations.
  {\sl Right panel}:
  The figure shows the $\gamma$-ray excess measured after the fitting and
  subtraction of $\gamma$-rays from the spatially extended source MGRO~J1908+06.
  The dashed box indicates the region shown in Fig.~\ref{fig:ss433_obs}.
  The jet termination regions e1, e2, e3, w1, and w2 observed in the X-ray data
  are indicated, as well as the location of the central binary. The dark lines
  show the contours of X-ray emission observed from this system.
  }
  \label{fig:ss433_region}
\end{figure}

A bright extended source, MGRO J1908+06, is detected with more than $30\sigma$
in this dataset and is located less than $2^\circ$ from the $\gamma$-ray
hotspots of SS~433 (Extended Data Fig.~\ref{fig:ss433_region}).  The region of
MGRO J1908+06 contains a pulsar and a supernova remnant, but it is not clear if
the observed TeV $\gamma$-ray emission is from either or even both of them. A
detailed discussion of MGRO J1908+06 is beyond the scope of this paper.
However, the morphology of MGRO J1908+06 must be carefully studied in order to
minimize the contamination of the emission due to MGRO J1908+06 on the fluxes
of the lobes.

A maximum likelihood analysis is performed that simultaneously fits the
emission from MGRO J1908+06 and the hotspots at w1 and e1. An electron
diffusion model appropriate for older pulsar wind
nebulae~\cite{Lopez-Coto:2017uku, Abeysekara:2017old} is used to describe the
spatial morphology of MGRO J1908+06. Given the uncertainty of the nature of
MGRO J1908+06, two other spatial models with Gaussian and power law radial
profiles are also tested in the simultaneous fit. The choice of spatial model
affects the best-fit fluxes from e1 and w1 at the level of $\pm20\%$. We adopt
this value as a systematic uncertainty on the flux from w1 and e1 due to VHE
emission from the nearby extended source.

\subsection{Contamination from Galactic Diffuse Emission}

The e1 and w1 regions are located roughly $2^\circ$ from the Galactic plane, so
the contamination from the Galactic diffuse emission (GDE) is negligible.
However, MGRO J1908+06 has a Galactic latitude of $\sim1^\circ$. Since the
three spatial models used to fit MGRO J1908+06 are radially symmetric, the
presence of GDE has the potential to produce an overestimate in the flux from
MGRO J1908+06, which could result in an underestimate in the flux measured from
w1 and e1. To minimize the effect of the GDE on the fit, the RoI is defined to
be a semicircular region centered on the position of MGRO J1908+06 (Extended
Data Fig.~\ref{fig:ss433_obs}). The RoI is designed to reduce the effect of GDE
by excluding the half of the source closest to the Galactic plane.

To estimate the systematic uncertainties associated with the choice of RoI and
possible contamination from GDE, a second maximum likelihood fit is performed
using a full-disk RoI which includes GDE, emission from MGRO J1908+06, and w1
and e1. The spatial distribution of the GDE is modeled with a
Gaussian profile of three different widths of $0.5^\circ$, $1^\circ$, and
$2^\circ$ in Galactic latitude, and is treated as constant in Galactic
longitude over the width of the RoI. Comparing the results to the fit with a
semicircular RoI indicates that contamination from GDE is less than 10\% for
e1 and less than 20\% for w1.

\subsection{Fit Results}

The fit results are reported in Extended Data Table~\ref{table:table_ss433}.
Fitting the emission from w1 and e1 simultaneously, we calculate
$\text{TS}=41.2$, which corresponds to a $5.4\sigma$ observation
($p=3.96\times10^{-8}$) after accounting for {\sl a posteriori} statistical
trial factors. To check the consistency of the results, we also fit w1 and e1
separately, re-defining the alternative model to include MGRO J1908+06 and only
e1 or w1. The significance of the VHE excess from these locations is below
$5\sigma$ in the fits, but the estimated fluxes are consistent with the
simultaneous fit.

An additional check is made on the effect of fixing versus floating the
best-fit positions of the emission from the east and west hotspots. In the
original alternative hypothesis, the point sources are centered on e1 and w1.
Here, the positions of the point sources are made additional free parameters in
the point source fit. We find that allowing the positions of the TeV hotspots
to vary does not affect the flux estimates, which are consistent with the
fixed-position fits. Moreover, the best-fit positions of the east and west TeV
emission regions are consistent with e1 and w1 within statistical
uncertainties.

\begin{table}[ht!]
  \refstepcounter{table}\label{table:table_ss433}
  {\bf Extended Data Table~\ref{table:table_ss433}~~Fits to the TeV emission from SS~433 using nested point source models}
  \begin{tabular}{cp{0.05\textwidth}p{0.20\textwidth}p{0.25\textwidth}p{0.10\textwidth}p{0.12\textwidth}}
    \hline
    & Lobe & Position\newline (RA, Dec)
       & $dN/dE$ at 20 TeV\newline [$10^{-16}$ TeV$^{-1}$cm$^{-2}$s$^{-1}$] & TS &
       Significance\newline(post trials)
    \\
    \hline
    \multicolumn{6}{l}{\bf Simultaneous fit to E+W hotspots, positions fixed.}
    \\
    & E & 19:13:37\newline 04$^\circ$55'48'' & $2.4^{+0.6+1.3}_{-0.5-1.3}$ & 41.2 & $5.4\sigma$
    \\
    & W & 19:10:37 \newline 05$^\circ$02'13'' & $2.1^{+0.6+1.2}_{-0.5-1.2}$ & &
    \\
    \multicolumn{6}{l}{\bf Separate fit to E+W hotspots, positions fixed.}
    \\
    & E & 19:13:37\newline 04$^\circ$55'48'' & $2.5^{+0.7+1.4}_{-0.5-1.4}$ &
    24.3 & $4.6\sigma$
    \\
    & W & 19:10:37 \newline 05$^\circ$02'13'' & $2.3^{+0.7+1.3}_{-0.5-1.3}$ &
    20.4 & $4.2\sigma$
    \\
    \multicolumn{6}{l}{\bf Separate fit to E+W hotspots, positions floated.}
    \\
    & E & 19:14:11$^{+20\text{s}}_{-39\text{s}}$\newline
          04$^\circ$59'10''$^{+03'30''}_{-06'18''}$ & $2.6^{+0.6+1.4}_{-0.5-1.4}$ &
    26.9 & $4.4\sigma$
    \\
    & W & 19:10:40$^{+17\text{s}}_{-17\text{s}}$ \newline
          05$^\circ$03'40''$^{+03'32''}_{-04'55''}$ & $2.4^{+0.6+1.3}_{-0.5-1.3}$ &
    23.4 & $4.0\sigma$
    \\
    \hline
  \end{tabular}
\end{table}


The choice of spectral model also affects the estimated $\gamma$-ray flux at
$20$~TeV. Extended Data Table~\ref{table:spectra} shows the dependence of the
best-fit VHE flux from e1 and w1 on the assumed spectral models, including
statistical uncertainties on the flux normalization at 20~TeV. Two spectral
models were tested: a simple power law $dN/dE_\gamma\propto
E_\gamma^{-\alpha}$, and a power law with an exponential cutoff
$dN/dE_\gamma\propto E_\gamma^{-\alpha}\exp{(-E_\gamma/E_\text{cut})}$. The
choice of spectral model can alter the flux normalization by almost a factor of
two compared to the default $E_\gamma^{-2}$ model.

\begin{table}[ht!]
  \refstepcounter{table}\label{table:spectra}
  {\bf Extended Data Table \ref{table:spectra}~~Dependence of Flux at 20~TeV on Spectral Assumption}
  \renewcommand{\arraystretch}{1.2}
  \begin{tabular}{ccccccc}
    \hline
   \multicolumn{7}{c}{\textbf{$dN/dE$ at 20 TeV [$\times 10^{-16}\,\textrm{TeV}^{-1}\,\textrm{cm}^{-2}\,\textrm{s}^{-1}$]}} \\
   & \multicolumn{2}{c}{Index: -1.5} & \multicolumn{2}{c}{Index: -2.0} & \multicolumn{2}{c}{Index: -2.5}
   \\
   Cutoff Energy & East Lobe & West Lobe & East Lobe & West Lobe & East Lobe & West Lobe \\
   \hline
    No Cutoff & $1.0^{+0.3}_{-0.2}$ & $0.9^{+0.3}_{-0.2}$ & $2.4^{+0.6}_{-0.5}$ & $2.1^{+0.6}_{-0.5}$ & $3.3^{+0.9}_{-0.7}$ & $2.4^{+0.9}_{-0.6}$
    \\[0.5em]
    $50$~TeV & $4.7^{+1.1}_{-0.9}$ & $4.2^{+1.1}_{-0.9}$ & $5.0^{+1.2}_{-1.0}$ & $4.1^{+1.3}_{-0.9}$ & $3.2^{+0.9}_{-0.7}$ & $1.7^{+1.1}_{-0.7}$
    \\[0.5em]
    $300$~TeV & $1.7^{+0.5}_{-0.4}$ & $1.6^{+0.5}_{-0.4}$ & $3.3^{+0.8}_{-0.7}$ & $2.9^{+0.8}_{-0.7}$ & $3.6^{+0.9}_{-0.7}$ & $2.4^{+0.9}_{-0.7}$ \\
    \hline
  \end{tabular}
  \renewcommand{\arraystretch}{1.0}
  \caption*{Fit results with different spectral models for the $\gamma$-ray emission.}
\end{table}

\subsection{Summary of Systematic Uncertainties}

The systematic uncertainties in the estimated fluxes from the TeV hotspots of
SS~433 include the following contributions: detector systematic effects,
modeling ambiguities in MGRO J1908+06, and contamination from Galactic diffuse
emission. The systematic uncertainties due to the modeling of MGRO J1908+06 and
the contamination from Galactic diffuse emission are $<\pm20\%$ and $-10\%$
($-20\%$) for the east (west) hotspot, respectively, and are discussed in
previous sections.

The detector response is estimated using Monte Carlo simulations and then
optimized using observations of the Crab Nebula~\cite{Abeysekara:2017mjj}, which
appears point-like in the HAWC data. Systematic uncertainties which potentially
affect the result presented here include the charge resolution and relative
quantum efficiency of the PMTs, the absolute quantum efficiency of the PMTs,
changes to the detector layout as construction proceeded, uncertainties in the
PSF, and systematic differences in the distribution of arrival times of
photoelectrons between data and simulation. The total systematic uncertainty on
the flux normalization from detector effects is $\pm50\%$.

All the components of the systematic uncertainties are summarized in Extended
Data Table~\ref{table:systematics} and combined in quadrature to estimate the
total systematic uncertainty on the VHE flux from the w1 and e1. We note that
since the systematic uncertainties due to MGRO~J1908+06 and GDE are
anti-correlated, the quadrature sum overestimates the total systematic
uncertainty. However, the effect is not particularly important, since the
detector systematic effects are the dominant source of uncertainty.

\begin{table}[ht!]
  \refstepcounter{table}\label{table:systematics}
  {\bf Extended Data Table \ref{table:systematics}~~Systematic uncertainties on the flux from SS~433}

  \begin{tabular}{lcc}
  \hline
  Systematic & East Lobe & West Lobe\\
  \hline
  Detector Systematic Effects & \multicolumn{2}{c}{$\pm 50\%$} \\
  MGRO J1908+06 Modeling & \multicolumn{2}{c}{$<\pm 20\%$} \\
  Galactic diffuse contamination & $-10\%$ & $-20\%$ \\
  \hline
  Total & $\pm 55\%$ & $\pm 55\%$ \\
  \hline
  \end{tabular}
  \caption*{
	Systematic $1\sigma$ error budget for the VHE $\gamma$-ray fits.}
\end{table}

\subsection{X-ray Template Fit and Upper Limit on the Extent of the Emission
Regions}

We performed several maximum likelihood fits modeling the hotspots as spatially
extended sources. In the first fit, we generated spatial templates for the
eastern and western regions based on the X-ray contours published by
ROSAT~\cite{Brinkmann:1996aa} and then performed a joint likelihood fit with
the two $\gamma$-ray hotspots and MGRO~J1908+06. This produces no improvement
in TS over a point-source fit.

In order to constrain the size of the $\gamma$-ray emission regions, likelihood
fits are applied using a Gaussian morphology convolved with the point spread
function of HAWC. To reduce the number of free parameters, we first fit
MGRO~J1908+06 using an RoI with SS~433 and its hotspots excluded. The extended
fit from MGRO~J1908+06 is then subtracted from the data, and the residual
$\gamma$-ray emission from the $\gamma$-ray hotspots is fit using two Gaussian
functions. The centers of the Gaussians are fixed to e1 and w1, and their
angular widths are estimated in a simultaneous fit to both the eastern and
western regions.

The maximum likelihood fit yields an angular width of $0.14^\circ\pm0.06$ for
the east hotspot and $0.08^{\circ}{}^{+0.14}_{-0.05}$ for the west hotspot. We
estimate the 90\% confidence region on the extent as the value of Gaussian
width which produces a decrease $\Delta\text{TS}=-2.71$ from the maximum
likelihood value. The resulting 90\% upper limits are $0.25^\circ$ for the east
region and $0.35^\circ$ for the west region.

\subsection{Upper Limit on Emission from the Central Binary}

In the present data set, no statistically significant emission is observed from
the center of SS~433. Using the method of Feldman and
Cousins~\cite{Feldman:1997qc}, we estimate the 90\% upper limit on the flux at
$20$~TeV to be $5.3\times10^{-17}$~TeV$^{-1}$~cm$^{-2}$~s$^{-1}$ after fitting MGRO J1908+06 and the emission at e1 and w1.

\subsection{Upper Limit on Detected $\gamma$-ray Energy}

The binning of $\gamma$-ray events into size bins $B$ causes us to lose
information about the energies of the $\gamma$ rays observed from SS~433. To
determine the upper energy bound on the flux we observe, we scan over the
maximum energy $E_{\gamma,\text{max}}$ used in the forward-folding analysis.
Starting at $E_{\gamma,\text{max}}=15$~TeV, we find that TS increases
monotonically until $E_{\gamma,\text{max}}=25$~TeV. Increasing
$E_{\gamma,\text{max}}$ above this value causes TS to plateau (for e1) or
decrease slightly (w1). We infer that the current measurement of e1 and w1
implies a minimum $E_{\gamma,\text{max}}=25$~TeV, and report this as a
conservative estimate of the highest energy observed by HAWC.

\subsection{Study of Residual Emission in the Region of Interest}

\begin{figure}[ht!]
  \centering
  \includegraphics[width=\textwidth]{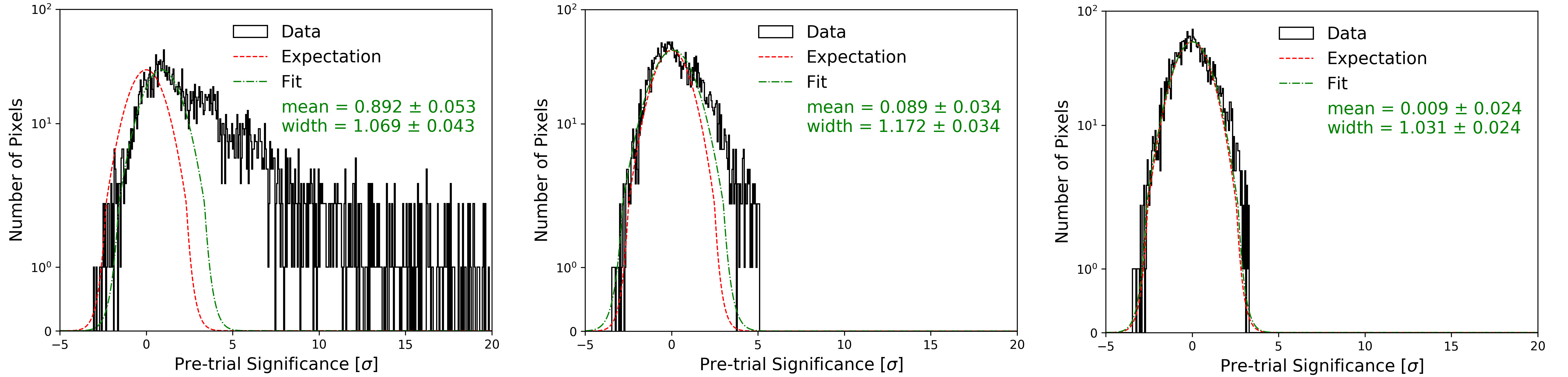}
  \caption{Distribution of pixel significance in the region of interest of the
  fit, defined as deviations from the background expectation, in the HAWC sky
  map (left), after fitting and subtraction of emission from MGRO~J1908+06
  (center), and after fitting and removal of emission from MGRO~J1908+06 and
  the $\gamma$ rays from w1 and e1 (right).}
  \label{fig:sighist}
\end{figure}

As a final check of the quality of the maximum likelihood fits, we plot the
distribution of the significance values in each HEALPix pixel in the RoI around
SS~433 in Extended Data Fig.~\ref{fig:sighist}. The significance values are
plotted in units of Gaussian $\sigma$. If only random background fluctuations
are present, the significance values follow a standard normal distribution,
shown by the dashed line in the figure.

Prior to the maximum likelihood fit, the significance distribution in the RoI
is considerably skewed toward positive values due to excess counts above
background from MGRO~J1908+06 and $\gamma$ rays from w1 and e1 (left panel of
Extended Data Fig.~\ref{fig:sighist}). After the subtraction of the maximum
likelihood fit to emission from MGRO~J1908+06 (center panel), the skew in the
distribution is considerably reduced, though still visible due to excess counts
from the interaction regions near SS~433. Finally, subtraction of the maximum
likelihood flux from w1 and e1 produces a distribution that, within statistical
uncertainties, is equivalent to background fluctuations (right panel).

\subsection{VHE Emission due to Leptonic Interactions}

In the leptonic scenario considered in this paper, relativistic electrons
scatter photons from the cosmic microwave background photons to TeV energies
via the inverse Compton process, and produce X-ray and radio emissions by the
synchrotron radiation. Although a far infrared background in the lobe may
contribute to the production of sub-TeV photons, no appreciable infrared
emission has been reported near the $\gamma$-ray emission region
\cite{Fabrika:2004asprv,Fuchs:2001asss}. The electron spectrum is obtained by
solving the continuity equation, considering radiative
cooling~\cite{Finke:2012wq}. The best-fit values of the parameters of the
injection spectrum, including the flux, spectral index, and maximum energy of
the electrons, and the magnetic field strength in the source region, are
obtained through Markov Chain Monte Carlo~\cite{ForemanMackey:2012ig} sampling
of their likelihood distributions when fitting to the multi-wavelength data.
The radio and soft X-ray data points correspond to the Effelsberg
Telescope~\cite{Geldzahler:1980aa} and the XMM-Newton (the Mos1
detector)~\cite{Brinkmann:2006zt} observations of a $6'$ circle centered on e1.
A 30\% uncertainty attributed to the unknown shape of the HAWC source is added
to the statistical and systematic errors of the observational data, though we
find the uncertainty has a negligible impact on the fit. The hard X-ray data
points and the sub-TeV upper limits are set by the RXTE, MAGIC, HESS, and
VERITAS observations of the e1 region~\cite{Safi-Harb:1999apj, Ahnen:2017tsc,
Kar:2017wvr}.

The VHE flux is determined using the flux from e1 at 20~TeV reported in
Extended Data Table~\ref{table:table_ss433}, where separate fits were made to
eastern and western hotspots, and the positions were fixed to e1 and w1.  The
best-fit values of the injection spectrum and magnetic field in the emission
region are $\alpha=1.87^{+0.04}_{-0.07}$,
$\log{(E_\text{max}/\text{PeV})}=3.53^{+0.31}_{-0.38}$, and
$B=16.04^{+2.60}_{-2.23}~\mu\text{G}$. Taking the distance to the source to
be 5.5~kpc, the fit suggests a total electron energy of $2.9\times10^{47}$~erg.
This is a small fraction of the total energy deposited by the jets of SS~433
over their lifetime, which is $\sim 9\times10^{50}$~erg assuming a kinetic jet
luminosity of $10^{39}$~erg~s$^{-1}$~\cite{Fabrika:2004asprv}. Future
multi-wavelength observations dedicated to the VHE $\gamma$-ray emission region
will better constrain the magnetic field strength and the properties of the
electron population.

We note that the presence of multi-hundred TeV to PeV electrons would challenge
the current particle acceleration mechanisms. Successful acceleration requires
that the acceleration rate $\dot{\gamma}_\text{acc}\approx eBv/m_ec^2$ based on
heuristic considerations (where $v$ is the velocity associated with the
notional electromotive force) exceed the cooling rate 
$
  \dot{\gamma}_\text{cool}\approx4c\sigma_T\gamma^2(B^2/8\pi)/3m_ec^2,
$
assuming that synchrotron radiation dominates the cooling processes in the
lobes of SS~433. This leads to a maximum electron energy
$
  E_{e,\text{max}}=271~\text{TeV}\ (v/100~\text{km~s$^{-1}$})^{1/2}\
                   (B/16~\mu\text{G})^{-1/2}.
$
For reference, the Alfv\'{e}n speed in the lobes is
$
  v_{A}=160~\text{km/s}\ (n_b/0.05~\text{cm}^{-3})^{-1/2}\
        (B/16~\mu\text{G}).
$
A higher Alfv\'{e}n speed could be achieved if the acceleration takes place in
the central spine of the jet, where the mass loading due to black hole
accretion is smaller and the magnetic field is stronger.
Depending on the exact electron acceleration mechanisms, $v$ could be
associated with the jet flow velocity, or with the Alfv\'{e}n speed. In both
cases, using these estimates, it is possible that the maximum electron energy
can exceed $1$~PeV. However, the timescale of acceleration mechanisms
such as second-order Fermi acceleration is proportional to $(v/c)^2$, making
the production of multi-hundred TeV electrons less efficient.
Future VHE $\gamma$-ray and hard X-ray observations can better constrain the 
electron cutoff energy, and diagnose the {\sl in situ} particle acceleration
mechanism.

\subsection{VHE Emission due to Hadronic Interactions}

\begin{figure}[ht!]
  \includegraphics[width=\textwidth]{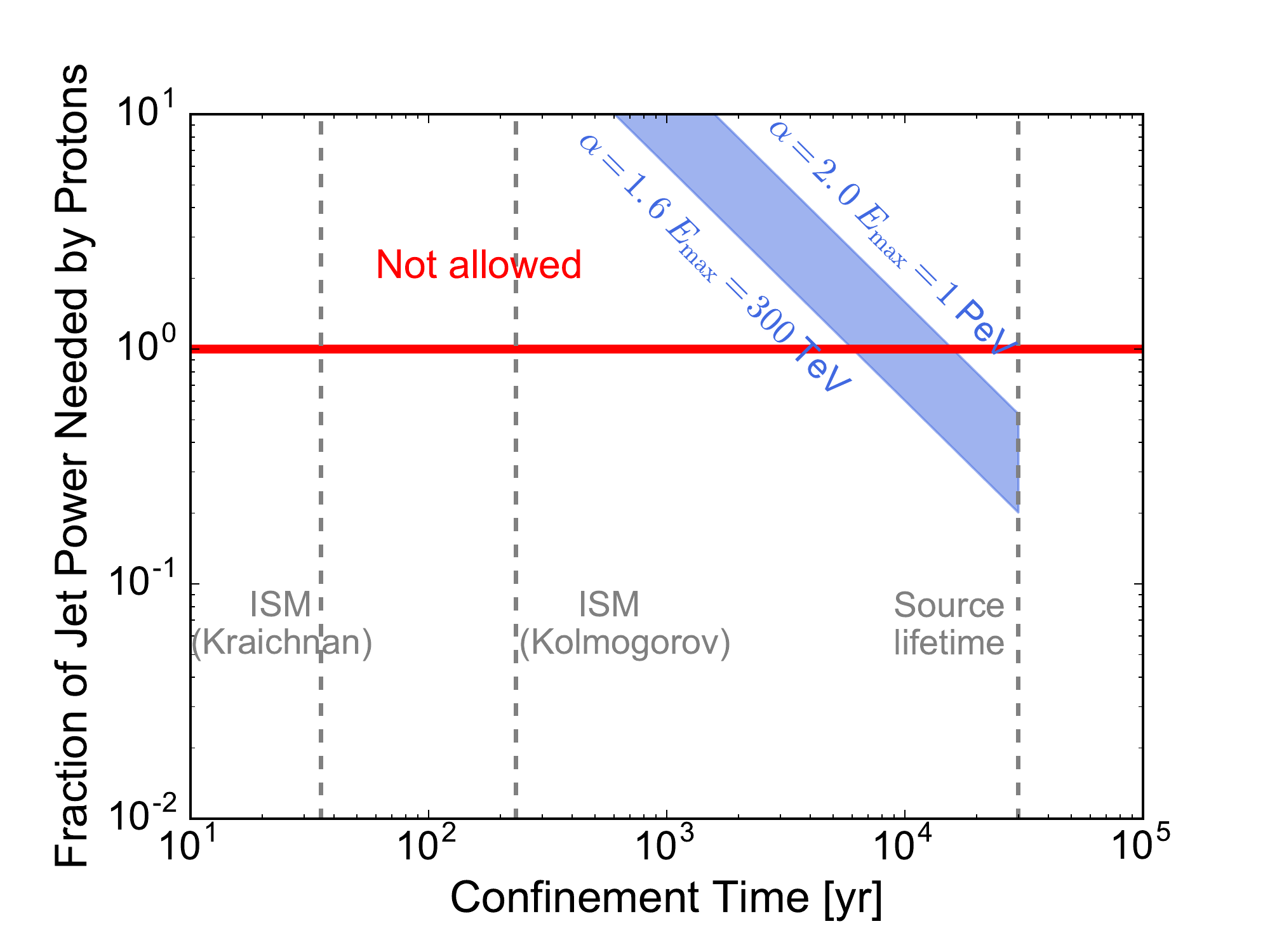}
  \caption{{\bf Fraction of jet power needed to produce the observed VHE gamma
  rays in the hadronic scenario. } The blue shaded region shows the energy
  injection rate of protons, in units of the kinetic luminosity of the jet, in
  order to produce the observed VHE $\gamma$ rays by interacting with ambient
  gas, as a function of the proton confinement time. A gas density of 
  $0.05\,\rm cm^{-3}$ is adopted for the source
  vicinity~\cite{Safi-Harb:1999apj, Panferov:2016rcw}.
  Most hadronic models require $>100\%$ jet power (above the red
  solid line) and are thus not allowed. Even when the diffusion coefficient is
  extremely small (for reference, the dashed grey lines show the source age and
  the confinement time of 200~TeV protons in a 30~pc region in the ISM with
  Kraichnan and Kolmogorov-type diffusion) and when the spectral index is much
  harder than 2, the hadronic scenario still requires a significant energy
  input from the jet.   }
  \label{fig1s:hadronic}
\end{figure}

In the hadronic scenario, high-energy protons interact with the ambient gas in
the source, and produce $\gamma$ rays via the decay
$\pi^0\rightarrow\gamma\gamma$.  Extended Data Fig.~\ref{fig1s:hadronic} shows
the fraction of jet power needed to be converted to protons to produce the
observed $\gamma$-ray flux. We assume a proton spectrum $dN/dE\propto
E^{-\alpha}\,\exp(-E/E_{\rm max})$, and adopt a proton-proton interaction cross
section of $\sim 50\,\rm mb$~\cite{Eidelman:2004wy}, and a baryon density of
$0.01-0.1\,\rm cm^{-3}$~\cite{Safi-Harb:1999apj, Panferov:2016rcw}. The total
proton energy is obtained by integrating this spectrum normalized to the VHE
$\gamma$-ray flux.

If the diffusion coefficient in the source is comparable to that in the ISM, no
hadronic models would be allowed, as they would require a proton injection rate
that exceeds the total kinetic luminosity of the jets of SS~433. Even in
extreme circumstances, e.g., where the diffusion coefficient is extremely small
(possibly due to scattering by turbulence generated from the streaming cosmic
rays \cite{Amato:2006vk, Malkov:2012qd}), particles could remain in the jet as
long as the jet lifetime of $10^4$~yr~\cite{Fabrika:2004asprv}. Assuming that
protons follow a hard spectrum with $\alpha<2$, the hadronic scenario would
still require that at least 30\% of the jet power goes to protons. While a
hadronic origin to the VHE flux is possible, it requires rather extreme source
parameters and is therefore disfavored.

{\bf References}
\vspace{1em}



\end{document}